\documentclass[11pt]{article}
\usepackage{Bedeschi_Moriond,epsfig}

\def\Journal#1#2#3#4{{#1} {\bf #2}, #3 (#4)}

\def\PRL{\em Phys. Rev. Lett.}

\def\be{\begin{equation}}
\def\ee{\end{equation}}
\def\bea{\begin{eqnarray}}
\def\eea{\end{eqnarray}}

\begin{document}
\vspace*{3.0cm}
\title{B LIFETIMES AND MIXING AT THE TEVATRON}

\author{ F. BEDESCHI ({\small for the CDF and D0 Collaborations}) }

\address{Istituto Nazionale di Fisica Nucleare, Sezione di Pisa,\\
Largo B. Pontecorvo, 3 Pisa, Italy}

\maketitle\abstracts{We present recent results on b-hadron lifetimes 
and mixing obtained from the analysis of the data collected at the Tevatron Collider
by the CDF and D0 Collaborations in the period 2002 - 2004. Many lifetime measurements have been
updated since the Summer 2004 conferences, sometimes improving significantly the
accuracy. Likewise the measurement of the $B_d$ oscillation
frequency has been updated. New limits on the $B_s$ 
oscillation frequency have been determined 
using for the first time Run II data.}

\section{Data samples}
\label{sec:Data}
The results described in this paper refer to the analysis of data collected at the Fermilab
Tevatron Collider from February 2002 till August 2004. 
During this period the accelerator has been continually
improving its luminosity delivering over 600 pb$^{-1}$ to each experiment.  
Initial store luminosities
exceeding $1\times10^{32} cm^{-2}sec^{-1}$ have now become quite common. 

The CDF and D0 detectors have been reliably taking data over this period. The data set sizes used for B 
physics analyses are in the range of 220 to 450 pb$^{-1}$.
	These data sets are collected with three major trigger categories:
\begin{itemize}
\item[-]{ di-lepton triggers: these provide a sample of millions of $J/\psi$'s many of which originate
	from the decay of a b-hadron. From this sample several thousands fully reconstructed
	b-hadrons are obtained;}
\item[-]{ single lepton triggers: these provide a large sample of b-hadrons decaying semi-leptonically. 
	The background is significantly reduced by requiring a fully reconstructed D meson with the
	right charge correlation with the lepton. Samples sizes of up to $\sim$ 100 thousand lepton plus D
	events are achieved;}
\item[-]{ triggers on displaced vertices: these provide large samples of b-hadrons decaying hadronically. 
	From these
	samples several thousand fully reconstructed hadronic B decays are obtained. 
	At present only CDF implements this kind of trigger.}
\end{itemize}

\section{b-hadron lifetimes}
\label{sec:Life}
The lifetime of b-hadrons is governed primarily by the decay of the b-quark, however contributions
from the spectator quarks can be up to $\sim$ 15\%. Presently these spectator effects are mostly calculated in
the framework of the Heavy Quark Expansion~\cite{HQE}. Theory errors on the ratio with the 
$B_d$ lifetime are in the few \% range as shown in table~\ref{tab:life}. The
experimental situation updated to the summer 2004 has been summarized in a 
recent report~\cite{HFAG05} of the Heavy Flavor Averaging Group (HFAG) and is also
 shown in table~\ref{tab:life}. 

\begin{table}[t]
\caption{HFAG summer 2004 averages
compared to theory calculations
\label{tab:life}.}
\begin{center}
\small
\begin{tabular}{|l|l|l|l|}\hline
b-hadron&Lifetime~\cite{HFAG05}&$\tau/\tau(B_d)$&$\tau/\tau(B_d)$\\
type&&experiments~\cite{HFAG05}&theory~\cite{HQE}\\ \hline
$B_d$&1.534 $\pm$ 0.013 ps&&\\ \hline
$B_u$&1.653 $\pm$ 0.014 ps&1.081 $\pm$ 0.015&1.06 $\pm$ 0.02\\ \hline
$B_s$&1.469 $\pm$ 0.059 ps&0.958 $\pm$ 0.039&1.00 $\pm$ 0.01\\ \hline
$\Lambda_B$&1.232 $\pm$ 0.072 ps&0.803 $\pm$  0.047&0.86 $\pm$ 0.05\\ \hline
\end{tabular}
\end{center}
\end{table}

\begin{figure}[!t]
\label{fig:D0_life}
\begin{center}
\epsfxsize 10cm
\epsffile{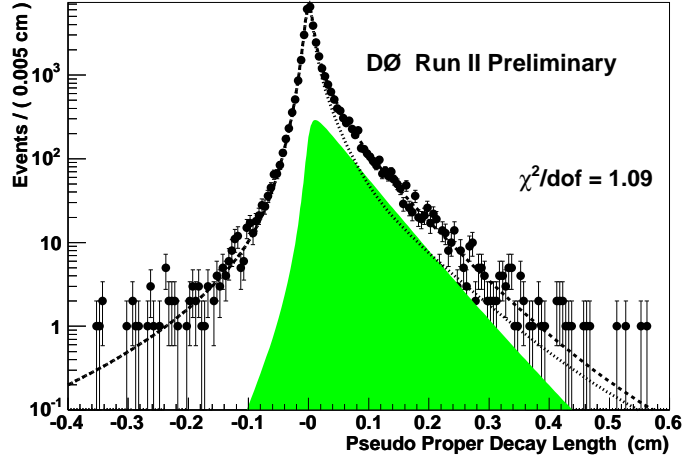}
\end{center}
\caption{D0 Collaboration: Pseudo-proper decay length distribution for $B_s^0$ semileptonic data 
with the result of the fit superimposed. The dotted curve represents the combinatorial background and the
filled histogram represents the $B_s^0$ signal.}
\end{figure}

Several new results which update the table above have been obtained by the CDF and D0 Collaborations. 
A summary of these new results is shown in table~\ref{tab:new}. Both 
experiments have improved their measurements with fully reconstructed modes involving a 
$J/\psi$~\cite{Bpsi}.
CDF has measured for the first time the lifetimes of $B$ mesons using completely reconstructed 
fully hadronic decays like $B\rightarrow D\pi$ or $B\rightarrow D3\pi$~\cite{CDFhad}; the small systematic 
error indicates a good control of the secondary vertex trigger efficiency turn on. CDF has also measured
the $B_u$ and $B_d$ lifetimes using a subsample of its semileptonic data~\cite{CDFsemi}.
D0 has measured the ratio of $B_u$ and $B_d$ lifetimes in their semileptonic samples with new technique
that involves fitting the bin by bin ratio of the lifetime distributions~\cite{D0ratio}.
Of particular interest is the D0 measurement of the $B_s$ lifetime in a very high statistics semileptonic 
sample~\cite{D0Bs}, as shown in fig.~\ref{fig:D0_life}. This is currently the best available measurement
of this quantity.

\begin{table}[!hb]
\caption{Summary of new lifetime results
\label{tab:new}}
\begin{center}
\tiny
\begin{tabular}{|l|l|l|l|l|l|l|}\hline
Luminosity&CDF $J/\psi$ modes&D0 $J/\psi$ modes&CDF hadronic&CDF semileptonic&D0 semileptonic&HFAG 2004\\
pb$^{-1}$&240&220&360&260&400&\\ \hline
$\tau(B_d)$ (psec)&1.539$\pm$0.051$\pm$0.008&1.473$\pm$0.051$\pm$0.023&1.511$\pm$0.023$\pm$0.013&
1.473$\pm$0.036$\pm$0.054&&1.534$\pm$0.013 \\ \hline
$\tau(B_u)$ (psec)&1.662$\pm$0.033$\pm$0.008&&1.661$\pm$0.027$\pm$0.013&1.653$\pm$0.029$\pm$0.032&
&1.653$\pm$0.014\\ \hline
$\tau(B_u)/\tau(B_d)$&1.08$\pm$0.042&&&1.123$\pm$0.040$\pm$0.040&1.08$\pm$0.016$\pm$0.014
&1.081$\pm$0.015\\ \hline
$\tau(B_s)$ (psec)&1.369$\pm$0.100$\pm$0.009&1.444$\pm$0.094$\pm$0.020&1.598$\pm$0.097$\pm$0.017&
&1.420$\pm$0.043$\pm$0.057&1.469$\pm$0.059 \\ \hline
$\tau(B_s)/\tau(B_d)$&0.890$\pm$0.072&0.980$\pm$0.073$\pm$0.003&&&&0.958$\pm$0.039\\ \hline
\end{tabular}
\end{center}
\end{table}

\section{$B_d$ and $B_s$ mixing}
\label{sec:Mix}

$B_d$ and $B_s$ mesons can turn into their anti-particle with a frequency which is related to the 
CKM matrix elements $V_{td}$ and $V_{ts}$ respectively. A measurement of both mixing frequencies, 
$\Delta m_d$ and $\Delta m_s$, would yield
a measurement of the ratio of these two elements with about 5\% theory uncertainty, thus providing
a strong constraint in global fits of the Unitary Triangle~\cite{UTfit}. 
Very precise measurements of $\Delta m_d$ have been available for some time, and are currently
dominated by the results of the B-factories~\cite{Bfact}. Recent CDF and D0 measurement~\cite{BdMix}
 are consistent with those results. 
$\Delta m_s$ is constrained to be larger than 14.5 ps$^{-1}$ at 95\% C. L. from previous 
work by the LEP experiments, SLD and CDF. Both CDF and D0 have recently obtained new limits on $\Delta m_s$. 
In the following we'll report on the CDF result, while the D0 result is described in a separate 
paper~\cite{D0mix}.

\begin{figure}[!b]
\label{fig:CDF_mix_peaks}
\begin{center}
\epsfysize 6cm
\epsffile{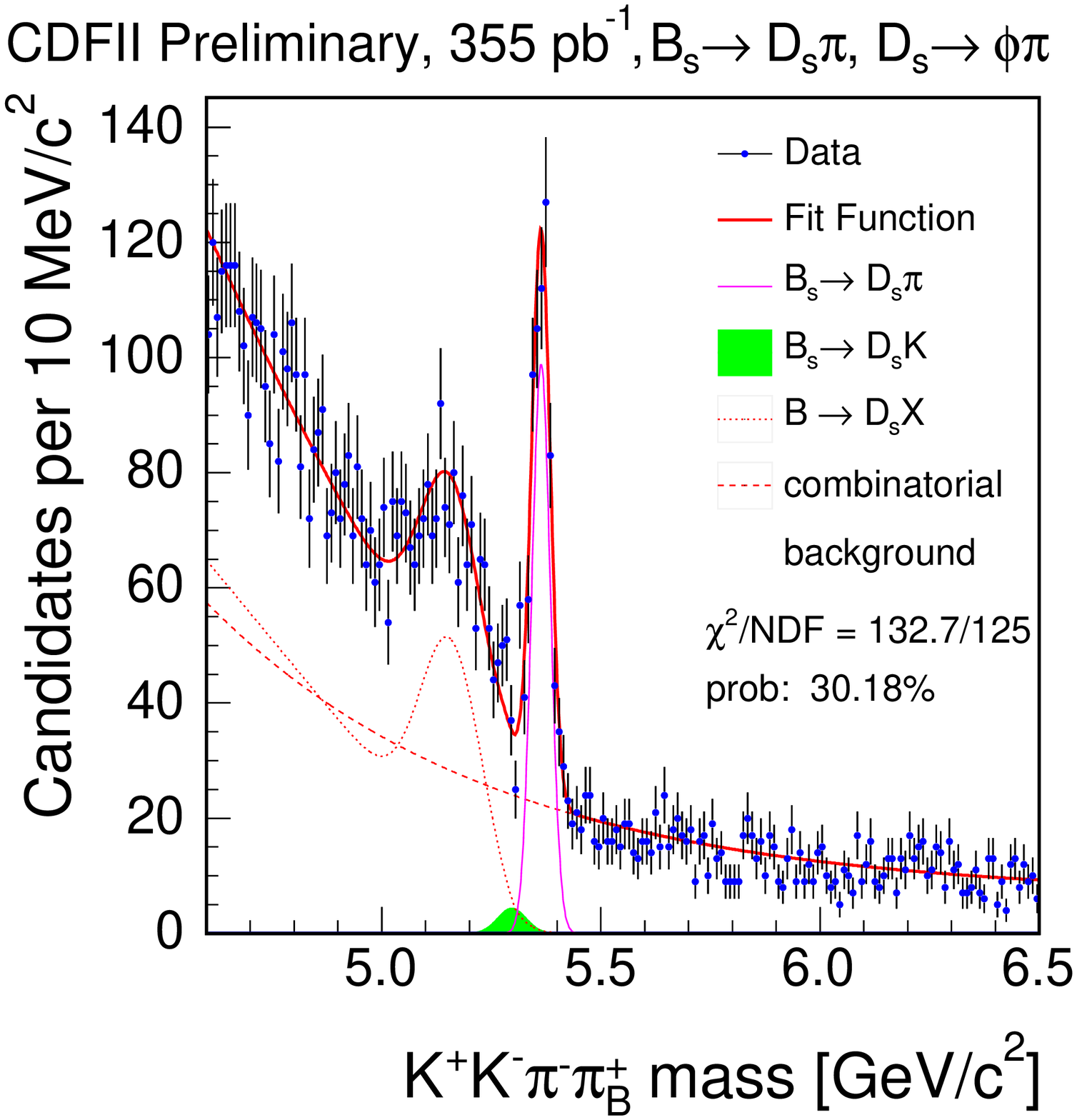}
\epsfysize 6cm
\epsffile{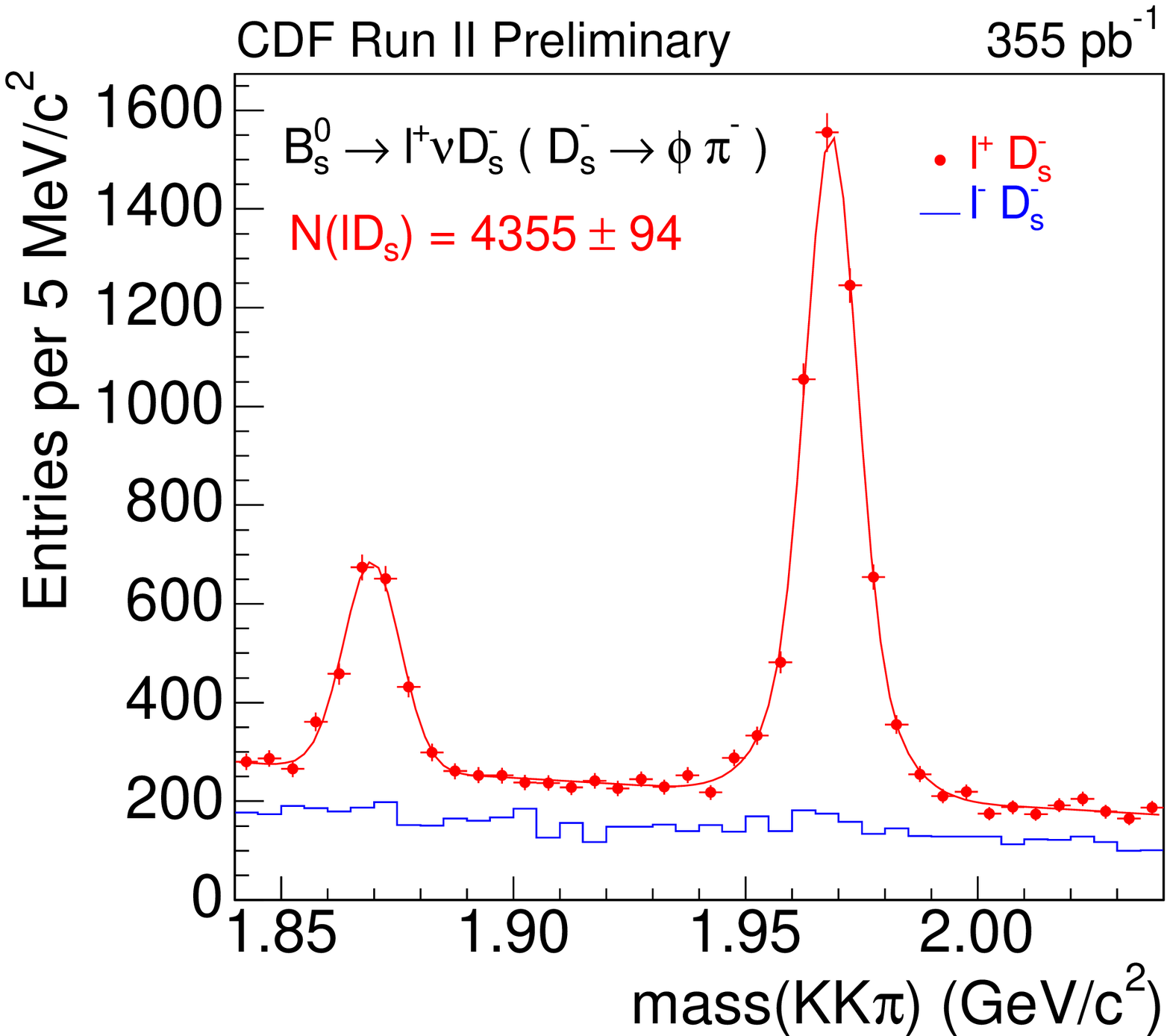}
\end{center}
\caption{CDF Collaboration: $B_s^0$ signals in the hadronic (left) and semileptonic (right) data samples.}
\end{figure}

CDF makes two parallel measurements~\cite{CDFmix} using both fully reconstructed
 hadronic decays and semileptonic 
decays with a fully reconstructed $D_s$ meson. In the former case about 900 $B_s\rightarrow D_s\pi$ 
events are found after summing over three possible $D$ decay modes: $\varphi\pi,\;K^*K$ and $\pi\pi\pi$;
the latter  yields a significantly larger statistics of about 7,500 $B_s\rightarrow l\nu D_s$ events,
however the proper time resolution is worse because of the incomplete knowledge of the decay kinematics
due to the missing neutrino.  Signals using the decay $D\rightarrow \varphi\pi$ are shown in 
fig.~\ref{fig:CDF_mix_peaks}.  

Flavor tagging is performed using only opposite side taggers. The tag sign is provided either by the
sign of an electron or muon, or by the average weighed charge of the tracks in a jet. A combined tagging 
power in the order of 1.4\% is obtained.

\begin{figure}[!t]
\label{fig:CDF_mix_scan}
\begin{center}
\epsfxsize 10cm
\epsffile{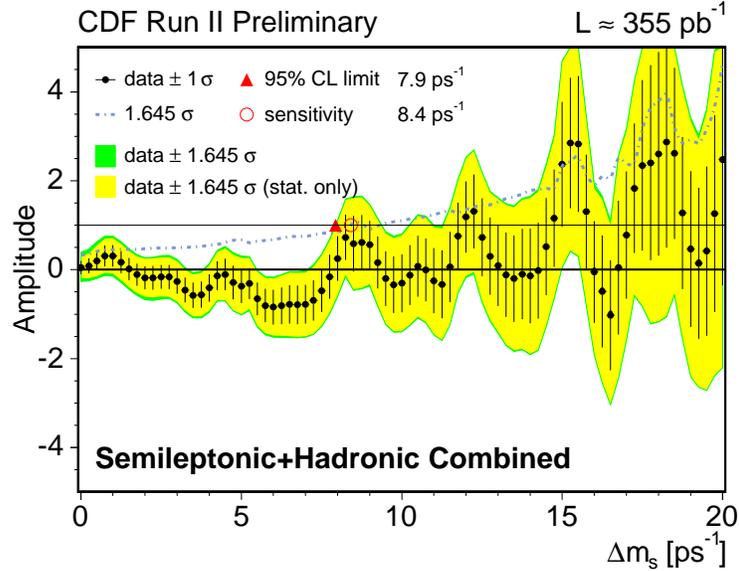}
\end{center}
\caption{CDF Collaboration: Amplitude scan combining hadronic and semileptonic $B_s^0$ signals.}
\end{figure}

CDF performs an amplitude scan~\cite{Roussarie} on both samples and then combines them to obtain a 95\% 
C.L. limit of 7.9 ps$^{-1}$ for an expected limit of 8.4 ps$^{-1}$. The scan result is shown 
in fig.~\ref{fig:CDF_mix_scan} and is clearly dominated by the statistical error. This result can be 
improved significantly in 
the near future by implementing more powerful taggers, improving the vertex resolution and including
more final states in this analysis. A positive observation will require more statistics, but appears to
be within reach.

\end{document}